\documentclass[journal=mamobx,manuscript=article]{achemso}

\usepackage{chemformula} % Formula subscripts using \ch{}
\usepackage[T1]{fontenc} % Use modern font encodings
\usepackage{graphicx}% Include figure files
\usepackage{dcolumn}% Align table columns on decimal point
\usepackage{bm}% bold math
\usepackage{natbib}
\usepackage{xcolor}

\author{C. Lang}
\email{c.lang@fz-juelich.de}
\affiliation{J\"ulich Centre for Neutron Science, Forschungszentrum J\"ulich, Germany}

\author{J. Kohlbrecher}
\email{joachim.kohlbrecher@psi.ch}
\affiliation{Laboratory of Neutron Scattering and Imaging, Paul Scherrer Institute, Villigen, Switzerland}

\author{L. Porcar}
\email{porcar@ill.fr}
\affiliation{Institut Laue-Langevin, Grenoble, France}

\author{A. Radulescu}
\email{a.radulescu@fz-juelich.de}
\affiliation{J\"ulich Centre for Neutron Science, Forschungszentrum J\"ulich, Germany}

\author{K. Sellinghoff}
\email{k.sellinghoff@fz-juelich.de}
\affiliation{Institute of Complex Systems 3, Forschungszentrum J\"ulich, Germany}

\author{J. K. G. Dhont}
\email{j.k.g.dhont@fz-juelich.de}
\affiliation{Institute of Complex Systems 3, Forschungszentrum J\"ulich, Germany}
\altaffiliation{Experimental Physics of Soft Matter, Heinrich Heine Universit\"at D\"usseldorf, Germany}

\author{M. P. Lettinga}
\email{p.lettinga@fz-juelich.de}
\affiliation{Institute of Complex Systems 3, Forschungszentrum J\"ulich, Germany}
\altaffiliation{Laboratory of Soft Matter and Biophysics, Katholieke Universiteit Leuven, Belgium}

\title{Microstructural understanding of the length and stiffness dependent shear thinning in semi-dilute colloidal rods}

\abbreviations{rheo-SANS, DEK}

\begin{document}

\begin{tocentry}
\includegraphics[width=8.5cm,height=3.5cm]{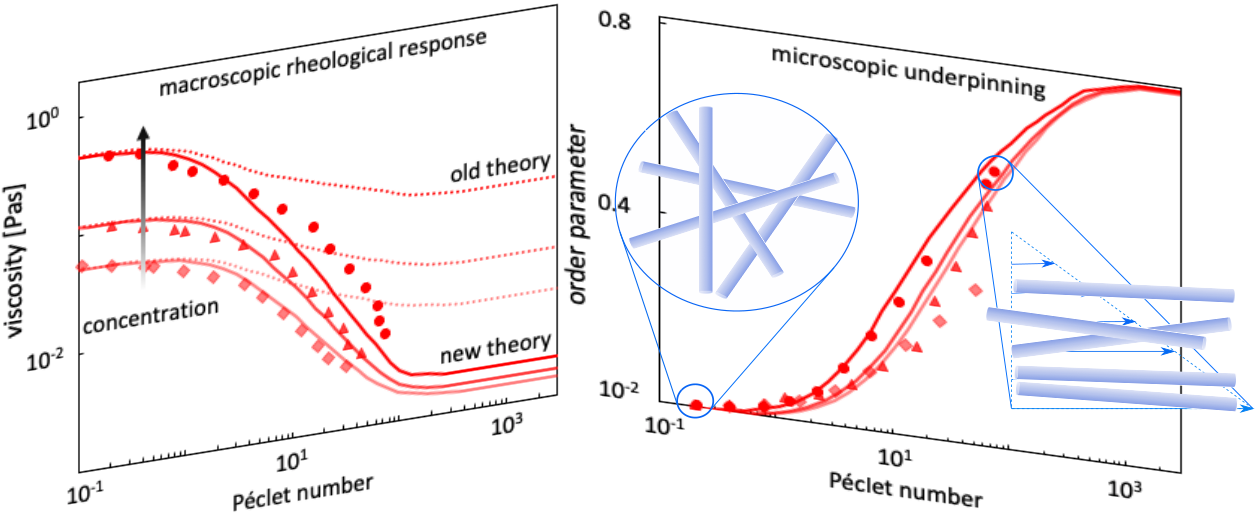}
\end{tocentry}

\begin{abstract}
  Complex fluids containing low concentrations of slender colloidal rods can display a high viscosity, while little flow is needed to thin the fluid. This feature makes slender rods essential constituents in industrial applications and biology. Though this behaviour strongly depends on the rod-length, so far no direct relation could be identified. 
  We employ a library of filamentous viruses to study the effect of rod size and flexibility on the zero-shear viscosity and shear-thinning behaviour. Rheology and small angle neutron scattering data are compared to a revised version of the standard theory for ideally stiff rods, which incorporates a complete shear-induced dilation of the confinement. 
  While the earlier predicted length-independent pre-factor of the restricted rotational diffusion coefficient is confirmed by varying the length and concentration of the rods, the revised theory correctly predicts the shear thinning behaviour as well as the underlying orientational order. These results can be directly applied to understand the manifold systems based on rod-like colloids and design new materials.
\end{abstract}

\section{Introduction}

Rod-like colloids are present in a broad diversity of daily-life products, like polymeric materials \cite{Wolfe1981,  McCulloch2006, Picken1990,Monteiro2015,Vigolo2000,Kouwer2013}, food-, and dairy products \cite{Tolstoguzov2000, Khan2007, vanderLinden2007, Araki2013}, and also play an important role in living organisms  \cite{Cherny2008, Olson2009, Maji2009, Stricker2010}. The reason is that at very low volume fractions rods will yield a fluid extremely viscous, while little flow is needed to thin the fluid \cite{Djalili2006, Boluk2011, Tanaka2014,Graf1993,Wierenga1998}. Though theory predicts that the mechanical response of fluids containing rod-like colloids is hugely affected by the particle length, flexibility and concentration, the lack of model systems has impeded a full understanding of the shear thinning behaviour.\par

The goal of this paper is to identify how the rod-like particle's aspect ratio, flexibility, and micro-structural order affect shear-thinning. To this end, it is important to develop a suitable model system  for which a controlled variation of aspect ratio and flexibility is feasible. In particular, it is important to have access to mono-disperse systems, since a variation in particle size and flexibility would blur the differences between the various systems. A very promising way of producing such mono-disperse systems is through  bio-engineering of rod-like viruses \cite{Dogic2016}. We use rod-like bacteriophages from the lambda class \cite{Sambrook1989}, namely fd wild type, fdY21M, M13k07, and Pf1 viruses, as those species span a range of $0.9-2\;\mu$m in length as well as a range of $1.2-9.9\;\mu$m in persistence length, while the thickness for all these viruses is close to roughly $7\;$nm. The effective rod thickness that determines the range of inter-rod interactions is modified by surface coating \cite{Grelet2016} and changing the ionic strength of the suspending medium \cite{Dogic2006}. These systems have proven to be ideally suited to study the flow behavior of rods\cite{Graf1993,Lettinga2005,Ripoll2008,Lang2016,Lang2019,Lang2019a}. It is equally important to perform experiments that can relate the flow induced orientational ordering with the rheological behaviour of the different systems. In-situ small angle neutron scattering in combination with rheology (rheo-SANS) is particularly suited to obtain this relation \cite{Foerster2005,Lang2016,Lang2019} because of the high contrast of this technique. Moreover, performing the scattering experiments yields the full 3D orientational distribution \cite{Lang2016,Liberatore2006,Liberatore2009}. Thus, the core of this paper is to determine the flow curves for all of the systems at varying concentration, measuring the viscosity and orientational ordering as a function of the applied shear rate with rheo-SANS. This allows us to relate the structural and mechanical response.\par 

The experimental results require an extension of the existing theory with a direct relation between the restricted motion of the rods and the degree of flow induced alignment. In order to understand this relation, we start from the observation that the excluded volume for other rods resulting from adding a slender rod to a fluid strongly increases with the concentration of the rods. This excluded volume causes a cascade of phase transitions with increasing concentration, depending on the rod aspect ratio \cite{Onsager1949,Grelet2008}, the flexibility \cite{Odijk1983, Odijk1986}, and the interactions between the rod-like particles \cite{ Vroege1992}. Similarly, these single-particle properties affect the flow behaviour in the disordered semi-dilute isotropic phase, where the number density $\nu=N/V_\text{tot}$ is already much larger than the overlap number density 
$\nu^\ast\sim1/L^3$ (with $N$ the number of particles, $L$ the particle length, and $V_\text{tot}$ the overall sample volume) \cite{Hess1976, Doi1981, Ranganathan1998, Morse1998a, Morse1998b, Morse1999, Switzer2003, Dhont2003}.
As a result, the rotational dynamics of particles is strongly reduced, leading to a strong shear-thinning behaviour \cite{Zirnsak1999,Hobbie2003, Lang2016}, where the viscosity strongly reduces with increasing shear rate, while the viscosity at zero-shear strongly increases with concentration. 
The reduced rotational diffusion is the base of the seminal theory for ideally stiff rods under shear flow, as proposed by Doi, Edwards, and Kuzuu \cite{Doi1978, Doi1981, Doi1986, Kuzuu1983}, 
where this aspect is taken into account. However, a full dilation of the confining tube in strong applied flow fields has not been realized. We will show how an extended version of the theory takes this dilation into account thus describing the complete  shear thinning behaviour in dispersions of colloidal rods.\par

This paper is structured as follows. We begin with a discussion of a theory for the shear-thinning behaviour and shear-induced nematic-like order in rod-like colloidal systems, followed by a description of the rheo-SANS procedures as well as the bio-engineered viruses used in this study. The experimental findings are detailed in section 
4, with a concluding discussion in section 5.

\section{Theory}
\label{2}

A theory for the response of suspensions of uni-axial, stiff, and mono-disperse rod-like particles to shear flow has been put forward by Doi, Edwards, and Kuzuu, which we will refer to as the Doi-Edwards-Kuzuu (DEK) theory \cite{Doi1978, Doi1981, Doi1986, Kuzuu1983}. A similar approach has been proposed by Hess \cite{Hess1976}. A microscopic foundation of the DEK theory is discussed in Ref.\cite{Dhont2003}, where the following  Fokker-Planck type equation of motion for the probability density function $\psi(\hat{\mathbf{u}};t)$ of the orientation $\hat{\mathbf{u}}$ of a test rod is derived from the $N$-particle Smoluchowski equation \cite{Smoluchowski1916}, 
\begin{equation}
\label{DB1}
\frac{\partial\psi}{\partial t}\,=\,D_{r}^{0}\,\hat{\mathcal{R}}\cdot \left[\hat{\mathcal{R}}\psi-\beta\psi{\bf\overline{T}}\right]-\hat{\mathcal{R}}\cdot\left[\psi \hat{\mathbf{u}}\times\left(\bf{\Gamma}\cdot \hat{\mathbf{u}}\right)\right]\;,
\end{equation}
where $D_{r}^{0}$ is the rotational diffusion coefficient of a freely rotating rod, $\bf{\Gamma}$ is the velocity-gradient tensor, and where
$\hat{\mathcal{R}}(\cdot)\,=\,\hat{\mathbf{u}}\times \nabla_{\hat{\mathbf{u}}}(\cdot)$ is the rotation operator, with $\nabla_{\hat{\mathbf{u}}}$ the gradient operator with respect to the Cartesian coordinates of $\hat{\mathbf{u}}$. Furthermore, $\overline{\mathbf{T}}$ is the effective torque acting on the test particle due to the presence of the other rods,
\begin{eqnarray}\label{DBtorque}
\!\!\!\!\!\!\overline{\mathbf{T}}=\!-\nu\!\!\int\!\! d\mathbf{R}\!\oint \!d\hat{\mathbf{u}}'\psi(\hat{\mathbf{u}}'\!;t)\,g(\mathbf{R},\hat{\mathbf{u}},\hat{\mathbf{u}}';t)\,\hat{\mathcal{R}}\mathcal{V}(\mathbf{R},\hat{\mathbf{u}},\hat{\mathbf{u}}')\,,
\end{eqnarray}
where $\nu$ is the number density of rods, $\mathbf{R}$ is the distance between the centres of two rods, $\mathcal{V}(\mathbf{R},\hat{\mathbf{u}},\hat{\mathbf{u}}')$ is the pair-interaction potential, and  $g(\mathbf{R},\hat{\mathbf{u}},\hat{\mathbf{u}}';t)$ is the pair-correlation function. The integral in eq.(\ref{DBtorque}) with respect to $\hat{\mathbf{u}}'$ ranges over the unit spherical surface. The torque in eq.(\ref{DBtorque}) is the torque $-\hat{\mathcal{R}}\mathcal{V}$ on the test rod due to interaction with a neighbouring rod, averaged with respect to the orientation and position of the latter. The effective torque is a function of concentration and shear rate through the pair-correlation function.\par

Within the DEK theory, instead of the effective torque in eq.(\ref{DBtorque}), the torque is assumed to take the form $-\hat{\mathcal{R}}\mathcal{V}_{\mbox{mf}}$, where $\mathcal{V}_{\mbox{mf}}$ is the mean-field Maier-Saupe potential $\sim\,\mathbf{S}\!:\!\hat{\mathbf{u}}\hat{\mathbf{u}}$, where $\mathbf{S}\,=\,<\hat{\mathbf{u}}\hat{\mathbf{u}}>$ is the orientational order parameter tensor. This mean-field potential can be derived from eq.(\ref{DBtorque}) by using the pair-correlation function $g=\exp\{-\beta\,\mathcal{V}\}$ for long and thin rods with (effective) hard-core interactions, which neglects the effect of shear flow on inter-rod correlations \cite{Dhont2003}. In the DEK-approach and its microscopic foundation in Ref.\cite{Dhont2003}, hydrodynamic interactions between the rods are neglected, which is a good approximation for long and thin rods.\par

The problem that the microscopic approach faces is that the analytical computation of an accurate expression for the pair-correlation function, including the effect of shear flow and multiple-rod interactions, is not feasible. In the approach in 
Ref.\cite{Dhont2003}, the pair-correlation function is simply approximated by the two-particle equilibrium pair-correlation function $g=\exp\{-\beta\,\mathcal{V}\}$ in 
eqs.(\ref{DB1},\ref{DBtorque}), leading to the following equation of motion for $\bf{S}$, which is valid up to third order in the orientational order parameter tensor,
\begin{eqnarray}
\label{DB0}
\frac{d\bf{S}}{dt}&=&-6 \,D_{r}^{0}\left[\mathbf{S}-\frac{1}{3}\mathbf{I}+\frac{L}{d}\,\varphi\left(\mathbf{S}^{(4)}:\mathbf{S}-\mathbf{S}\cdot \mathbf{S}\right)\right] \nonumber \\
&&\;\;\;\;\,+\dot{\gamma}\,\left[\,\hat{\mathbf{\Gamma}}\cdot \mathbf{S}+\mathbf{S}\cdot\hat{\mathbf{\Gamma}}^T-2\mathbf{S}^{(4)}:\hat{\mathbf{E}}\,\right]\;,
\end{eqnarray}
where $\dot{\gamma}$ is the shear rate, $\bf{I}$ is the identity tensor, $\hat{\bf{\Gamma}}=\bf{\Gamma}/\dot{\gamma}$, and similarly $\hat{\bf{E}}=\bf{E}/\dot{\gamma}$ with $\mathbf{E}$ the symmetric part of the velocity-gradient tensor, $L$ is the length and $d$ the diameter of the rods, and $\varphi=\nu Ld^2\pi/4$ is the volume fraction of rods. Furthermore, $\bf{S}^{(4)}=<\hat{\mathbf{u}}\hat{\mathbf{u}}\hat{\mathbf{u}}\hat{\mathbf{u}}>$ for which a closure relation will be discussed below. This approximation accounts only for simultaneous interactions of just two rods, which is, as shown by Onsager \cite{Onsager1933, Onsager1949}, exact for very long and thin rods in equilibrium up to concentrations well within the nematic state. Under flow conditions, however, inter-rod correlations are affected by dynamical events which involve topological constraints where multiple-rod interactions are essential. Like in the DEK theory, we will account for such topological constraints by replacing the bare rotational diffusion coefficient $D_{r}^{0}$ by an effective diffusion coefficient $D_{r}^{\text{eff}}$ that accounts for such constraints. On first sight one might think that $D_{r}^{0}$ should only be replaced by the effective diffusion coefficient in the term $\sim (L/d)\,\varphi$ in eq.(\ref{DB0}) that involves the pair-correlation function. This leads, however, to a thermodynamic inconsistency\cite{TD_inconsist}. 
We therefore replace the bare diffusion coefficient in eq.(\ref{DB0}) also for the ideal relaxation contribution $\sim (\mathbf{S}-(1/3)\hat{\bf{I}})$. The interpretation for such an overall replacement is that each rod is embedded in an effective medium, which also affects the otherwise ideal relaxation contribution. Replacing the bare diffusion coefficient $D_{r}^{0}$ by the effective diffusion coefficient $D_{r}^{\text{eff}}$ leads to the following dimensionless equation of motion,
\begin{eqnarray}
\label{DB}
\frac{d\bf{S}}{d\tau}&=&-\,\left(\mathbf{S}-\frac{1}{3}\mathbf{I}\right)-\frac{L}{d}\varphi\left(\mathbf{S}^{(4)}:\mathbf{S}-\mathbf{S}\cdot \mathbf{S}\right) \nonumber \\
&&\;\,+\mbox{\small{$\frac{1}{6}$}}\,Pe\,\left[\,\hat{\mathbf{\Gamma}}\cdot \mathbf{S}+\mathbf{S}\cdot\hat{\mathbf{\Gamma}}^T-2\mathbf{S}^{(4)}:\hat{\bf{E}}\,\right]\;,
\end{eqnarray}
where the dimensionless time $\tau = 6\,D_{r}^{\text{eff}}\,t$ is introduced, and where the dimensionless effective rotational Peclet number is given by,
\begin{eqnarray}\label{peclet}
Pe\,=\,\dot{\gamma}\,/D_{r}^{\text{eff}}\;.
\end{eqnarray}
This is the fundamental equation of motion on the basis of which the experiments concerning shear-induced alignment will be discussed. To this end, however, the higher order average $\mathbf{S}^{(4)}$ must be expressed in terms of $\bf{S}$. Within the DEK theory, a simple quadratic factorization closure relation is used. Here, we will employ the more accurate closure relation that has been proposed in Ref.\cite{Dhont2003},
\begin{eqnarray}\label{closu}
\mathbf{S}^{(4)}:\mathbf{M}\,=\,\mbox{\small{$\frac{1}{5}$}}\left\{\mathbf{S}\cdot\mathbf{\overline{M}}+\mathbf{\overline{M}}\cdot\mathbf{S}-\mathbf{S}\cdot\mathbf{S}\cdot\mathbf{\overline{M}}-\mathbf{\overline{M}}\cdot\mathbf{S}\cdot\mathbf{S}+2\,\mathbf{S}\cdot\mathbf{\overline{M}}\cdot\mathbf{S}+3\,\mathbf{S}\,\mathbf{S}:\mathbf{\overline{M}}\right\}\,,
\end{eqnarray}
where,
\begin{eqnarray}\label{closu2}
\bf{\overline{M}}\;=\;\mbox{\small{$\frac{1}{2}$}}\,\left[\bf{M}+\bf{M}^{T}\right]\;,
\end{eqnarray}
is the symmetric part of the tensor $\bf{M}$. In the construction of this closure relation, the quantity $\bf{S}^{(4)}:\bf{M}$ is assumed to be a linear combination of all possible terms that are linear or  second order in $\bf{S}$. The coefficients of these terms in such a linear combination are then determined from the requirements that the closure should be valid in the isotropic and perfectly aligned state, for which $\bf{S}^{(4)}$ is known, and from the fact that $\bf{S}^{(4)}:\bf{M}$ is symmetric and it's trace is equal to $\bf{S}:\bf{M}$. Note that various more accurate closure schemes exist for different non-equilibrium situations \cite{Kroeger2008}.

Within the realm of a tube model for rotational diffusion in simple shear flow, Doi proposed an expression for the effective diffusion coefficient which is valid in the semi-dilute regime and for moderate degrees of rod-alignment \cite{Doi1978}. In the present paper we present experiments up to very high shear rates, where there is a high degree of alignment. For such high shear rates, the tube model no longer applies, and the rotational diffusion coefficient is essentially equal to $D_{r}^{0}$. We thus need to interpolate between Doi's expression and $D_{r}^{0}$. Such an interpolation is established by the following expression, 
\begin{eqnarray}\label{rotDiff4}
D_{r}^{\text{eff}}\,=\frac{D_{r}^{0}}{1+\frac{1}{c}\,\left[\frac{5}{4}\,\nu\,L^{3}\,\left(\,1-\frac{3}{5}\,\mathbf{S}:\mathbf{S}-\frac{2}{5}\,(\mathbf{S}:\mathbf{S})^{2}\,\right)\,\right]^{2}}\;.
\end{eqnarray} 
This expression essentially coincides with the original Doi expression in the semi-dilute regime and for moderate shear rates, and becomes equal to $D_{r}^{0}$ for low concentrations\cite{Takada1991} as well as high shear rates. The coefficient $c\approx 1.3\times 10^3$ was obtained by Teraoka et al. \cite{Teraoka1985} and by Tao et al.\cite{Tao2006} using Brownian dynamics simulations. The concentration dependence of the effective diffusion coefficient in the absence of shear flow found in Ref.\cite{Tao2006} is in agreement with 
eq.(\ref{rotDiff4}) over the entire concentration range. In section 4 we discuss how $c$ can be determined experimentally.\par

Within the approach in Ref.\cite{Dhont2003}, the relation between the deviatoric part of the stress tensor $\mathbf{\Sigma}_{D}$ and the orientational order parameter tensor is based on a microscopic expression that is valid for rigid colloids as derived in Ref.\cite{Dhont2002}. Using  $D_{r}^{0}=3\,k_{B}T\ln\{L/d\}/\pi \eta_{s}\,L^{3}$, we find 
\begin{eqnarray}
\label{eStress1}
\bf{\Sigma}_{D}&=&2\eta_{s}\dot{\gamma}\hat{\mathbf{E}}+3\nu k_{B}T\left\{\left[\mathbf{S}-\frac{1}{3}\mathbf{I}+\frac{L}{d}\varphi\left(\mathbf{S}^{(4)}:\mathbf{S}-\mathbf{S}\cdot \bf{S}\right)\right]\right. \nonumber\\
&&\left.\;\;\;\;\;\;\;\;\;\;\;\;\;\;\;\;\;\;\;\;\;\;\;\;\;\;\;\;\;\;\;\;\;\;+\frac{1}{6}Pe^{0}\left[\mathbf{S}^{(4)}:\hat{\mathbf{E}}-\frac{1}{3}\mathbf{I}\,\bf{S}:\hat{\mathbf{E}}\right]\right\}\,,
\end{eqnarray}
where $\eta_{s}$ is the shear viscosity of the solvent and
\begin{eqnarray}\label{Pebare}
Pe^{0}\;=\;\dot{\gamma}/D_{r}^{0}\;,
\end{eqnarray}
is the bare Peclet number.

Note that the effective Peclet number does not appear in this expression.  Although the pre-factor of the last term in eq.(\ref{eStress1}) contains the functional form of the inverse bare diffusion coefficient, it cannot be interpreted in terms of a diffusion coefficient, also because it is not related to the only contribution $\sim (L/d)\,\varphi$ that is connected to the pair-correlation function. The stress in the present approach is therefore only indirectly affected by entanglements through the dependence of $\bf{S}$ on the effective diffusion coefficient.
A similar expression for the stress tensor, based on thermodynamic arguments, has been put forward within the DEK-theory. \par

The zero-shear viscosity can be obtained from eqs.(\ref{DB},\ref{eStress1}) by expanding 
$\mathbf{S}$ up to first order in the shear rate. From the stationary equation of motion (\ref{DB}) it is found that,
\begin{eqnarray}\label{zero1}
\mathbf{S}\;=\;\frac{1}{3}\,\hat{\mathbf{I}}+\frac{1}{15}\,\frac{\dot{\gamma}}{\tilde{D}_{r}^{\text{eff}}}\,\hat{\mathbf{E}}+\mathcal{O}(Pe^{2})\;,
\end{eqnarray}
where a second effective diffusion $\tilde{D}_{r}^{\text{eff}}$ is introduced, that also accounts for the slowing down of orientational diffusion due to the proximity of the isotropic-nematic spinodal, which is given by,
\begin{eqnarray}\label{effspinod}
\tilde{D}_{r}^{\text{eff}}\,=\,\left(1-\,\frac{L}{5d}\,\varphi\right) D_{r}^{\text{eff}}\;.
\end{eqnarray}
Substitution into eq.(\ref{eStress1}), and again linearisation with respect to the shear rate, it is readily found that the zero-shear viscosity $\eta_{0}$ is given by\cite{Doi1978},
\begin{eqnarray}\label{zero2}
\eta_{0}\;=\;\eta_{s}+\nu\,k_{B}\,T\left[\frac{1}{30}\,\frac{1}{D_{r}^{0}}+\frac{1}{10}\,
\frac{1}{D_{r}^{\text{eff}}}\right]\;.
\end{eqnarray} 
The first term in the square brackets is a viscous contribution that arises from the expression (\ref{eStress1}) for the stress tensor, while the second contribution is an elastic contribution, as it arises from correlations between the rods as captured by the equation of motion (\ref{DB}) for the order parameter tensor. In the semi-dilute regime and for the small shear rate under consideration here, eq.(\ref{rotDiff4}) 
reduces to the original Doi's expression. Substitution of $\mathbf{S}=(1/3)\hat{\mathbf{I}}$, the appropriate expression for the rotational diffusion coefficient that is to be used in the above expression for the zero-shear viscosity is therefore $D_{r}^{\text{eff}}=c\,D_{r}^{0}\,/\nu^{2}\,L^{6}$. In the semi-dilute regime, where $D_{r}^{0}\gg D_{r}^{\text{eff}}$, the first contribution in eq.(\ref{zero2}) is of minor importance.   

Equations (\ref{DB},\ref{closu},\ref{rotDiff4},\ref{eStress1},\ref{zero2}) present a full theoretical prediction for the concentration- and shear-rate dependent orientational order and viscosity for stiff rods that will be used in section \ref{4} in comparison to experimental results. \par    

\section{Experiments and Materials}
\label{3}

\subsection{Experiments}
\label{3.1}

Rheo-SANS studies were carried out at the SINQ spallation source in Villigen, Switzerland, the Heinz Maier-Leibnitz Zentrum in Garching, Germany, and the Institut Laue-Langevin in Grenoble, France. An Anton Paar MCR 501 rheometer (Anton Paar, Graz, Austria) was mounted in the SANS-1 and KWS-2 neutron beams for simultaneous measurements of shear stress and orientational ordering in the velocity-vorticity plane. The used sample environment was a Couette cell with a gap size of $1\,$mm, and with an inner-cell radius of $48\,$mm. Since the Anton Paar MCR 501 is a stress-controlled rheometer, all rheological measurements were repeated with a strain-controlled ARES LS rheometer (TA Instruments, New Castle, PA, USA). For measurements in the velocity-gradient plane, a shear cell was mounted at the D22 diffractometer. For measurements at extremely low shear rates, of the order $\dot\gamma\sim 10^{-5}\,$s$^{-1}$, the shear cell was equipped with a state-of-the-art brushless ec-motor and a gear box transducer. All measurements were performed at $25^{o}$C with a detector distance of $4.6\,$m and a wavelength of $0.6\pm0.1\,$nm at SANS-1, a detector distance of $5.6\,$m, a wavelength of $0.6\pm 0.1\,$nm, and an aperture size of $3$x$0.15~$mm at D22, and a detector distance of $8\,$m and a wavelength of $0.5\pm0.1\,$nm at KWS-2. From the available $q$-range at the different instruments, we selected a small subset of $3.2$x$10^{-2}$ to $4.6$x$10^{-2}\,$~\AA~(in the Porod regime) for our analysis. In this regime, the different reduction schemes for extracting the orientational distribution of rods from two-dimensional scattering patterns by Hayter and Penfold \cite{Hayter1984} and Maier and Saupe \cite{Maier1959}  give similar results. \par

\begin{figure}[h!]
	\begin{center}
		\includegraphics[scale=0.55]{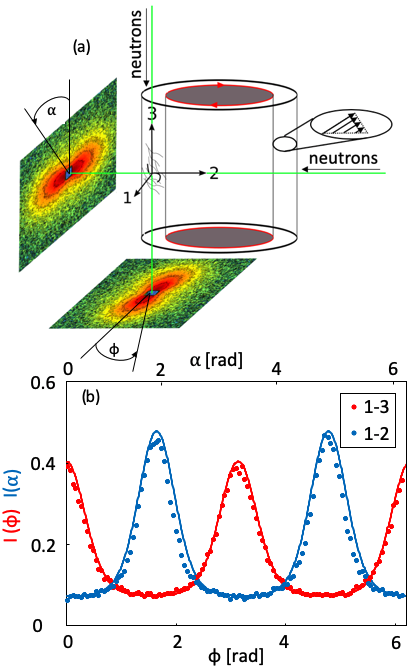}
		\caption{(a) Sketch of the rheo-SANS set-up (Couette cell). (b) Intensity measured in the velocity-vorticity (1-3) and velocity-gradient (1-2) plane as a function of the azimuthal angle for fdY21M at a concentration of 10.6~mg/ml with $\dot\gamma=64~\text{s}^{-1}$. Lines are fitted Maier-Saupe distributions.}
		\label{fig0}
	\end{center}
\end{figure}

We find the projections of the orientational distribution in the flow-vorticity (1-3) plane from the intensity profiles as a function of $\phi$, and in the flow-gradient (1-2) plane from the intensity profiles as a function of $\alpha$, see FIG.~\ref{fig0}, and by fitting these profiles with a Maier-Saupe type \cite{Maier1959} of orientational distribution function, $f(\phi)=I_0\exp(\tilde{c}P_2(\phi-\phi_\text{max})-1)$.
Here $I_0$ is the amplitude, $\tilde{c}$ the width, and $\phi_\text{max}$ the tilt angle of the projected nematic director with respect to the direction of flow. From this distribution, we find the projected order parameters as $<\!P_{2}\!>=\int_0^1d\!\cos(\phi)f(\phi)P_2(\phi-\phi_\text{max})/\int_0^1d\!\cos(\phi)f(\phi)$. The identical relations can be written for $\phi$ replaced by the angle $\alpha$. From these projections, the largest eigenvalues, $\lambda_1(\phi)$, and $\lambda_1(\alpha)$, of the projections of the traceless orientation tensor $\tilde{\mathbf{Q}}$ are found as $\lambda_1(\phi)=(2\langle P_2(\phi)\rangle+1)/3$ (and identical for $\alpha$), and since the tilt angle in the vorticity-gradient direction is effectively 0,  we find the traceless orientation tensor in the reference frame of the measurement as $\tilde{Q}_{33}=\lambda_1(\phi)$, $\tilde{Q}_{22}=\mathcal{T}-\lambda_1(\phi)/2$, and $\tilde{Q}_{33}=-\mathcal{T}-\lambda_1(\phi)$, where $\mathcal{T}=(\lambda_1(\phi)\lambda_1(\alpha)-\lambda_1(\phi)^2)/2(2-\lambda_1(\phi)-\lambda_1(\alpha))$. From those quantities, the full tensor $\tilde{\mathbf{S}}=(2\tilde{\mathbf{Q}}+\mathbf{I})/3$ can thus be constructed and rotated into the microscopic reference frame via $\mathbf{S}=\left(\mathbf{R}^{(0,\phi)}\right)^T\tilde{\mathbf{S}}\mathbf{R}^{(0,\phi)}$, where $R^{(0,\phi)}_{11}=R^{(0,\phi)}_{22}=\cos\theta$, $R^{(0,\phi)}_{12}=-R^{(0,\phi)}_{21}=\sin\theta$, $R^{(0,\phi)}_{33}=1$ and all other elements of the rotation tensor are zero. \cite{Lang2016} Finally, we obtain the actual order parameter $<\!P_{2}\!>$ from the largest eigenvalue of the full tensor $\mathbf{S}$.

\subsection{Materials}
\label{3.2}

The rod-like bacteriophages fd wild type, fdY21M, M13k07 and M13-mini were grown in their host-bacteria in Luria-Bertani broth, following standard biological protocols \cite{Sambrook1989}. FdY21M virus is a stiff mutant of wild-type fd virus. M13k07 is a derivative of the M13 wild type bacteriophage. All of these phages grow in E.-coli of the pBluescript KS(-) type, except for  Pf1 virus, which is a Pseudomonas Aeruginosa phage purchased from Asla Biotech.
\par 

After purification by ultra-centrifugation, a fraction of fdY21M was suspended in Phosphate buffer and coated with end-functionalized mono-disperse 8~kDa Polyethylen Glycol (peg) in a grafting-to procedure based on click-chemistry \cite{Zhang2010}. The functionalized material was carefully cleansed from buffer and peg residuals by repeated ultra-centrifugation and re-dispersion. \par

The peg-coated fdY21M as well as bare fdY21M and all other viruses were suspended in $20\,$mM Trizma base buffer solutions of Deuterium Oxide with $90\,$mM Sodium Chloride, corresponding to an ionic strength of $100\,$mM at $pH\, 8.3$. One fraction of bare fdY21M was re-dispersed in a $20\,$mM Trizma base buffer solution of Deuterium Oxide, resulting in $10\,$mM ionic strength at the same $pH$ value. A change in ionic strength of the buffer corresponds to a change in effective thickness of the rods. We are using effective thicknesses here, which have been determined experimentally for fd virus \cite{Dogic2006}. For $100\,$mM, an effective thickness of $d_\text{eff}=10.5$~nm is found which increases for $10\,$mM to $d_\text{eff}=17$~nm, which is equal to the experimentally determined thickness of the peg-coated virus. Table~\ref{tab1} gives an overview of the length, $L$, and the persistence length, $L_\text{p}$, of the various viruses. \par

\begin{table}[h!]
	\caption{Length and persistence length of the different viruses.}
	\smallskip{}
	\centering{}
	\begin{tabular}{ccc}
		\hline \hline
		\textbf{material} & \textbf{length} & $L_\text{p}$ \cite{Barry2009}  \\
		& ($\mu$m) & ($\mu$m) \\
		\hline
		fd & 0.88 & 2.8$\pm$0.7 \\
		fdY21M & 0.91 & 9.9$\pm$1.6  \\
		M13k07 & 1.2 & 2.8$\pm$0.7 \\
		%M13-mini & 330 & 2800$\pm$700 & 100 & 10.5 \\
		Pf1 & 2.1 & 2.8$\pm$0.7 \\
		\hline \hline
	\end{tabular}
	\label{tab1}
\end{table}

All colloidal suspensions were first prepared at a  concentration just below the isotropic-nematic coexistence region, and subsequently diluted to concentrations of roughly $75,\, 50,\, 25$ and $10\%$ of the lower-binodal concentration. These concentrations are well within the semi-dilute regime \cite{Graf1993}.

\section{Results and Discussion}
\label{4}

\begin{figure}[tbp]
	\begin{center}
		\includegraphics[scale=0.5]{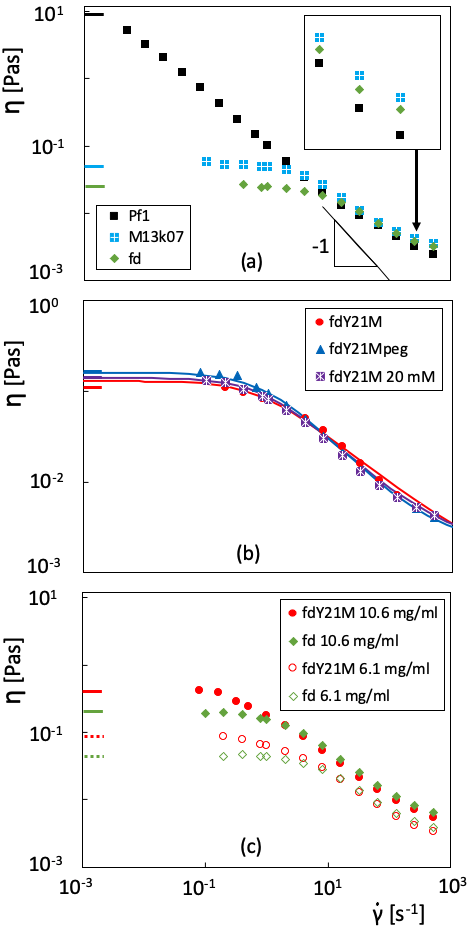}
		\caption{(a) Viscosity curves of Pf1 ($L=1.96~\mu$m), M13k07 ($L=1.2~\mu$m), and fd ($L=0.88~\mu$m) at fixed persistence length of $L_p=2.8\pm0.7~\mu$m, and concentration of 4.8~mg/ml. Inset: (magnified) viscosity at high shear rates. (b) Viscosity curves of fdY21M with different thicknesses at a concentration of 6.8~mg/ml. The lines are Carreau fits with params. (top to bottom) $a=\{1.1,1.1,1.4\}$~sec., $b=\{0.7,0.65,0.54\}$, and $\eta_s=8.9$x$10^{-4}$~Pas. (c) Viscosity curves of fd wild type and fdY21M for two concentrations. The short lines at the vertical axis display the zero-shear viscosities obtained from scaling, see e.g. FIG.~\ref{fig2}.}
		\label{fig1}
	\end{center}
\end{figure}

In FIG.~\ref{fig1} we plot flow curves, that is, the viscosity as a function of shear rate, for suspensions where we vary only the contour length (FIG.~\ref{fig1}(a)), only the thickness (FIG.~\ref{fig1}(b)), and only the persistence length (FIG.~\ref{fig1}(c)). We compare data at fixed weight concentrations as the mesh size of a network of stiff rods, $\xi$, or tube diameter, is identical for the different particles: $\xi$ can be estimated as $\xi=L\sqrt{\nu^{*}/\nu}$, where $\nu^{*}$ is the overlap number density \cite{deGennes1976}. Since $\nu^{*}\sim L^{-3}$ and  $\nu\sim w[\text{mg/ml}]/L$ (as the molecular weight is proportional to $L$), it follows that $\xi\sim \sqrt{w[\text{mg/ml}]}$. For a weight concentration of $w[\text{mg/ml}]=4.8\,$mg/ml the mesh size is equal to $0.11 \,\mu$m, which is much smaller than the length of the viruses, but is large compared to the diameter of the rods.\par

FIG.~\ref{fig1}(a) shows the flow curves for rods with varying contour length but identical persistence length. The viscosity at low shear rates strongly decreases in the order of decreasing length: an increase in length by a factor of 2 between fd virus and Pf1 causes an increase in the zero-shear viscosity by more than a decade.
For the viscosities at high shear rates the trend is partly reversed, as can be seen from the inset in FIG.~\ref{fig1}(a): the flexible Pf1 now exhibits the lowest viscosity, while the other viruses retain their length-order dependence. This could be due to the fact that rods with $L_p/L\approx3$ can form hairpins located within planes perpendicular to the gradient direction, which strongly reduce friction, as was shown for F-actin strands in Refs.\cite{Kirchenbuechler2014, Huber2014}. At high shear rates there is thus a mix of shear-alignment and anisotropic disentanglement that determines the viscosity.\par

FIG.~\ref{fig1}(b) shows the influence of thickness on the rheological behaviour for fdY21M, at a concentration of $6.8\,$mg/ml. The thickness of the virus is varied from $7$ to $17\,$nm by changing the ionic strength of the suspending medium ($20$ and $100\,$mM) and by coating the virus with peg. As can be seen from this figure, all three viscosity curves coincide. Therefore, the thickness of the rod does not play a role, neither for small nor for large shear rates. \par

FIG.~\ref{fig1}(c) shows the influence of the flexibility. Here, the viscosity is plotted for fd (with $L_{p}/L=3.2$) and fdY21M (with $L_{p}/L=10.9$) at two different concentrations ($6.1$ and $10.6\,$mg/ml) in the semi-dilute concentration regime. Clearly, an increased flexibility of particles significantly lowers the zero-shear viscosity. The length dependence of the viscosity is, however, much more pronounced than the flexibility dependence. \par
The slope of the viscosity curve of the longest rod in this study, Pf1, in the low shear-rate regime is significantly steeper than that of the other systems (see Fig.~\ref{fig1}(a)). For a slope of $d\log\eta/d\log\dot\gamma\approx-1$ one would expect to find flow instabilities. In the low- to intermediate shear-rate region, Pf1 has a slope close to but not quite as steep as $-1$, while all other systems display slopes of $-0.5$ or less. In addition, the stiffness also leads to a stronger shear thinning (see Fig.~\ref{fig1}(c)). This points in the direction that a further increase in particle length as well as their stiffness could provoke flow instabilities such as those observed e.g. for xanthane gum \cite{Tang2018}. It is known, however, that none of the systems under study here show flow-instabilities in the long-time regime \cite{Lang2019}.\par
\begin{figure}[tbp]
	\begin{center}
		\includegraphics[scale=0.5]{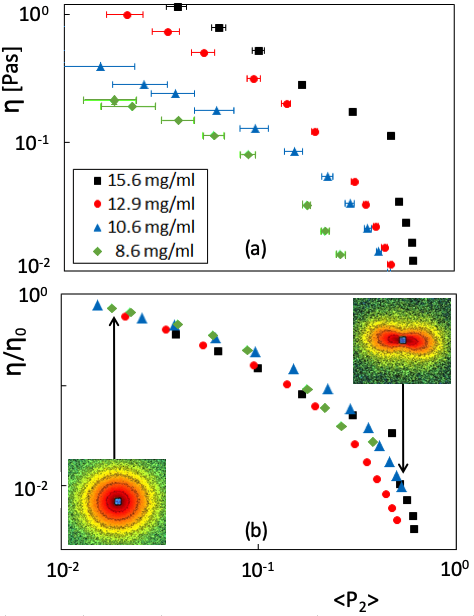}
		\caption{(a) Viscosity and (b) relative viscosity versus order parameter for different concentrations of fdY21M ($L=0.91~\mu$m, $L_p=9.9\pm1.6~\mu$m). Error bars for the viscosity are inside the symbols. Insets in (b) display scattering patterns for 10.6 mg/ml at low and high shear rates.}
		\label{fig2}
	\end{center}
\end{figure}

Accurate values for the zero-shear viscosities $\lim_{\dot{\gamma}\to0}\eta=\eta_0$ of most samples can be obtained by a fit of the flow curve to the empirical Carreau equation \cite{Carreau1972}, $(\eta-\eta_{s})/(\eta_{0}-\eta_{s})=[1+a\dot{\gamma}^{2}]^{-b}$, where $\eta_{s}$ is the viscosity of the solvent, $\eta_{0}$ is the zero-shear viscosity, and the parameters $a$ and $b$ are fitting variables, see examples in FIG.~\ref{fig1}(b). This fitting method, however, becomes ambiguous in the case of the very long rod, Pf1, where there is no clear sign of a zero-shear plateau. Therefore, we additionally employ a scaling scheme introduced by F\"orster et al. \cite{Foerster2005}, where the viscosity is plotted as a function of the scalar orientational order parameter instead of the shear rate. FIG.~\ref{fig2}(a) shows the unscaled viscosities as a function of the order parameter for different concentrations of fdY21M in the semi-dilute concentration regime. FIG.~\ref{fig2}(b) shows that the different concentrations approximately super impose if the viscosity is scaled with an apparent zero-shear viscosity. From this scaling procedure, we obtain a second estimate for $\eta_0$, which compares well with the values from the Carreau fits and the viscosity close to the plateau region, see the short lines at the abscissas of FIG.~\ref{fig1} and the direct comparison of the scaling and the Carreau fit in FIG.~\ref{fig1}(b). \par

\begin{figure}[tbp]
	\begin{center}
		\includegraphics[scale=0.55]{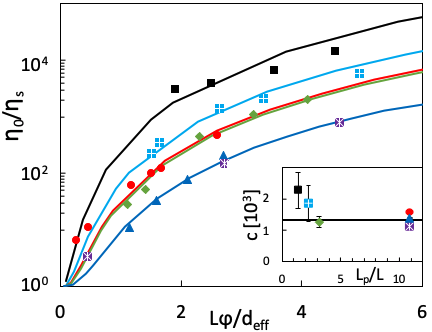}
		\caption{zero-shear viscosity divided by the solvent viscosity as a function of the scaled volume fraction of rods. The markers are identical with those of FIG.~\ref{fig1}. Lines represent the theoretical prediction, equation~\ref{zero2}. Inset: The fitted values of the constant $C$ in Doi's expression for the diffusion coefficient as a function of the particle stiffness. The horizontal line is the estimate from Ref.~\cite{Teraoka1985}.}
		\label{fig3}
	\end{center}
\end{figure}

Combining the two techniques, we obtain accurate values for the zero-shear viscosities of all samples. The zero-shear viscosities for various viruses are plotted in FIG.~\ref{fig3} as a function of concentration. The lines correspond to the prediction in eq.(\ref{zero2}), where the factor $c$ in the expression for the effective diffusion coefficient has been used as a fitting parameter. The values for the fitting parameter $c$ in Doi's expression for the effective diffusion coefficient (see the discussion just below eq.~\ref{zero2}) are shown in the inset in FIG.~\ref{fig3} as a function of $L_{p}/L$. With the exception of Pf1 (with the smallest value of $L_{p}/L=1.33$), the values for $c$ are independent of the flexibility to within experimental error. The values of $c$ varies between $1.1$x$10^{3}$ and $2.3$x$10^{3}$. This is comparable to the pre-factor $c=1.3$x$10^{3}$ found from computer simulations in Refs.\cite{Teraoka1985,Tao2006}, which value corresponds to the horizontal line in the inset in FIG.~\ref{fig3}, and a factor three lower than what was found for more flexible polymeric systems\cite{Sato1991,Petekidis2000}. That the zero-shear viscosities can be reasonably well described by expressions for stiff rods, including the quite flexible virus Pf1, shows that the zero-shear viscosity is relatively insensitive to flexibility as compared to the rod-length dependence.

\begin{figure}[tbp]
	\begin{center}
		\includegraphics[scale=0.6]{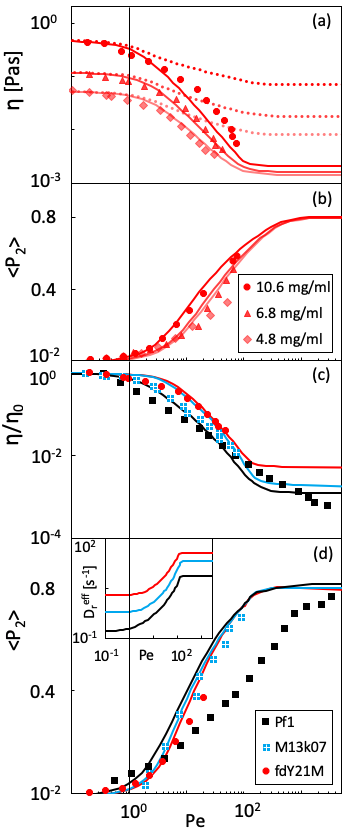}
		\caption{(a) Viscosity and (b) order parameter as a function of $Pe$ for fdY21M at different concentrations. (c) Viscosity scaled by the zero-shear viscosity and (d) order parameter as a function of the  Peclet number for three rods with different lengths, Pf1 ($L=1.96~\mu$m, $L_p=2.8\pm0.7~\mu$m), M13k07 ($L=1.2~\mu$m, $L_p=2.8\pm0.7~\mu$m), and fdY21M ($L=0.91~\mu$m, $L_p=9.9\pm1.6~\mu$m) for a common concentration of $\varphi=0.024$. The thick lines are theoretical predictions using eqs.~\ref{rotDiff4}, \ref{DB}, and 
			\ref{eStress1}, dashed lines in (a) combine eqs.~\ref{DB}, and \ref{eStress1} with Doi's original diffusion coefficient. The vertical line marks $Pe=1$. Inset: Rotational diffusion, equation~\ref{rotDiff4}, defining the Peclet number scaling.}
		\label{fig4}
	\end{center}
\end{figure}

Having established the value of $c$ in eq. \ref{rotDiff4}, we can proceed to describe the full shear thinning behaviour, without any further fitting parameters, using 
eq.~\ref{eStress1}, and the underlying nematic-like ordering quantified by the full order parameter $\langle P_2\rangle$ (as obtained from the full tensor $\mathbf{S}$),  using eq.~\ref{DB}. Results are shown in FIG.~\ref{fig4} (a,b), for the ideal stiff rod, fdY21M, at varying concentration and in FIG.~\ref{fig4} (c,d) for rods with varying length and fixed concentration. Here the shear rate is always scaled by the effective orientation dependent rotational diffusion, see eq. \ref{rotDiff4}, yielding the relevant Peclet number, $Pe=\dot\gamma/~D_r^\text{eff}$. That this is the proper scaling is confirmed by the fact that for all systems shear thinning sets in around $Pe=1$, where shear forces start to dominate rotational Brownian forces. 

For the ideal, stiff rod we see that the shear rate dependence as well as the concentration dependence of both $\eta$, FIG.~\ref{fig4}(a), and  $\langle P_2\rangle$ , FIG.~\ref{fig4}(b), are well described by our revised theory over a very large range of shear rates. The agreement for Peclet numbers beyond unity, the good agreement is due to the effect of tube-dilation, which is accounted for by the proposed interpolation between Doi's expression for the rotational diffusion coefficient and the free diffusion coefficient in eq.~\ref{rotDiff4}. With increasing orientational ordering, the space available for rotation increases strongly, reaching the point of full rotatability where $D_r^\text{eff}=D_r^0$, see the inset of FIG.~\ref{fig4}(d). Of course, the rods will not fully explore the created available space due to the high values of $Pe$. However, if we do not allow for this strong tube-dilation in our theory, the prediction of the high shear-rate viscosity deviates considerably for the measured values, see dotted lines in FIG.~\ref{fig4}(a). Due to the limited shear rate that can be reached with a conventional rheometer we could not observe the levelling off in ordering and viscosity for this system.

The length dependence of the shear thinning process is also well captured, as can be seen in FIG.~\ref{fig4}(c,d), where we find a master-plot when scaling $\eta$ by $\eta_0$. However, contrary to the relatively stiff rods, there is a significant deviation between theory and experiment in  the shear rate dependent orientational ordering and shear thinning for the flexible Pf1, as can be seen in FIG.~\ref{fig4}(d). Ordering and thinning set in at lower shear rates and are less pronounced at high shear rates. A comparatively low ratio between $L_p$ and $L$ seems to lead to a comparatively modest increase in orientational ordering as compared to systems with a lower flexibility in terms of $L_p/L$. The reason could be hairpin formation, as discussed above. Moreover, the effect of hydrodynamic friction, as suggested by Hinch and Leal \cite{Hinch1972}, should not be ignored, as shown previously \cite{Lang2019a}. We have discussed this issue in an earlier paper, where we treated the linear dynamic response as well as extensional flow response of fd and fdY21M. Interestingly, Morse theory for the linear response\cite{Morse1998a,Morse1998b}, which includes flexibility, did not give as good a correspondence as the non-linear theory which does not include flexibility. 
In its simplest form, the effective diffusion could be adapted\cite{Sato1991b} in the $N$-particle Smoluchowski equation, while in a more sophisticated version, the connection between slightly flexible rods and ideal polymer chains should be made, possibly along the lines of the work by Frey et al. \cite{Frey1998}, and Broedersz and MacKintosh \cite{Broedersz2014}. We leave this for future studies.

\section{Conclusions}
\label{5}

A library of filamentous viruses probed by rheo-SANS experiments was used to asses the effect of flexibility and length of the rods on their rheological behaviour in the semi-dilute concentration regime. The zero-shear viscosity as well as the shear-rate dependence of the viruses are contrasted with a  microscopic theory based on the $N$-particle Smoluchowski equation \cite{Dhont2003}, resulting in very similar expressions for the order parameter and stress tensor as derived earlier by Doi, Edwards, and Kuzuu \cite{Doi1978, Doi1981, Doi1986, Kuzuu1983} (here referred to as the DEK theory). Similar to the DEK theory, dynamical correlations in the equation of motion for the order parameter are accounted for by replacing the single-rod rotational diffusion coefficient with an effective diffusion coefficient \cite{Doi1978,Kuzuu1983}. Contrary to their theory, our theory includes the full shear-rate (as well as concentration) dependence of the tube diameter, achieving full dilation at high Peclet numbers. 
Together with an expression of the suspension stress in terms of the orientational ordering tensor, the presented theoretical framework describes the zero-shear viscosity, shear-thinning, and the underlying nematic ordering of the ideal stiff rods without any fit parameter, see Fig.\ref{fig4}.

The zero-shear viscosity is found to increase with increasing length of the rod-like viruses, and decreases with increasing flexibility as can be seen in Fig.~\ref{fig2}. The length dependence of the shear viscosity is, however, much more pronounced as compared to its dependence on flexibility, especially at shear rates beyond the shear rate where shear thinning sets in. As a result, the zero-shear viscosity, and in particular the shear-rate dependence of the shear viscosity, as well as the orientational order, can be reasonably described by the theory developed for ideally stiff rods even for quite flexible rods, provided that $L_{p}/L$ is larger than about $1.7$. With our combined analytical and experimental effort, we can confirm the pre-factor $c\approx 1.3$x$10^3$, as proposed by computer simulations for Doi's rotational diffusion coefficient of crowded rod suspensions in the semi-dilute concentration regime \cite{Teraoka1985,Tao2006}.

These findings together with our improved theoretical treatment allow for analytical predictions of flow properties for a large variety of systems based on rod-like particles. As such, they bear potential for use in industrial applications, where an a priori knowledge of structure-property relations e.g. in the production process is highly beneficial, and for understanding biological processes, as stiff filaments are essential building blocks for cells.\par

\begin{acknowledgement}

This research is funded by the European Union within the Horizon 2020 project under the DiStruc Marie Sk\l{}odowska Curie innovative training network; grant agreement no. 641839.

\end{acknowledgement}


%\bibliography{bibGeom}
\providecommand{\latin}[1]{#1}
\makeatletter
\providecommand{\doi}
  {\begingroup\let\do\@makeother\dospecials
  \catcode`\{=1 \catcode`\}=2 \doi@aux}
\providecommand{\doi@aux}[1]{\endgroup\texttt{#1}}
\makeatother
\providecommand*\mcitethebibliography{\thebibliography}
\csname @ifundefined\endcsname{endmcitethebibliography}
  {\let\endmcitethebibliography\endthebibliography}{}

\end{document}